\documentclass[usenatbib]{mn2e}
\usepackage{natbib}
\usepackage{apjfonts}
\usepackage{amsmath}

\usepackage{epsfig}
\usepackage{graphicx}
\usepackage{amssymb}
\usepackage{aas_macros}
\usepackage{color}

\newcommand{\kms}{\,{\rm km \, s^{-1}}}
\newcommand{\pc}{\,{\rm pc}}
\newcommand{\kpc}{\,{\rm kpc}}

\newcommand{\oversim}[2]{\protect{\mbox{\lower0.5ex\vbox{%
   \baselineskip=0pt\lineskip=0.2ex
   \ialign{$\mathsurround=0pt #1\hfil##\hfil$\crcr#2\crcr\sim\crcr}}}}} 

\catcode`\"=\active\let"=\"  

\def\3{{\ss} }

\def\c12{{1\over 2}}

\def\d{{\rm d}}   
   
\def\plusplus{\raise 0.3ex\hbox{${\scriptstyle ++}$}{}}

\def\and{{{\rm M}31}}

\def\gyr{\,{\rm Gyr}}
\def\myr{\,{\rm Myr}}
   
\bibliography{biblio}

\begin{document}   
\title[Wide binaries in ultra-faint dSphs]{Wide binaries in ultra-faint galaxies: a window onto dark matter on the smallest scales}
\author[Jorge Pe\~{n}arrubia et al.]{Jorge Pe\~{n}arrubia$^{1}$\thanks{jorpega@roe.ac.uk}, Aaron D. Ludlow$^{2}$, Julio Chanam\'e$^{3,4}$, Matthew G. Walker$^5$\\
$^1$Institute for Astronomy, University of Edinburgh, Royal Observatory, Blackford Hill, Edinburgh EH9 3HJ, UK\\
$^2$Institute for Computational Cosmology, Dept. of Physics, University of Durham, South Road, Durham DH1 3LE, UK\\
$^3$Instituto de Astrof\'isica, Pontificia Universidad Cat\'olica de Chile, Av. Vicu\~na Mackenna 4860, 782-0436 Macul, Santiago, Chile\\
$^4$Millennium Institute of Astrophysics, Santiago, Chile.\\
$^5$McWilliams Center for Cosmology, Department of Physics, 5000 Forbes Ave., Carnegie Mellon University, Pittsburgh, PA 15213, USA
}
\maketitle  

\begin{abstract} 
We carry out controlled $N$-body simulations that follow the dynamical
evolution of binary stars in the dark matter (DM) haloes of
ultra-faint dwarf spheroidals (dSphs). We find that wide binaries 
with semi-major axes $a\gtrsim a_t$ tend to be quickly disrupted by the tidal
field of the halo. In smooth potentials the truncation scale, $a_t$, is mainly governed by 
(i) the mass enclosed within the dwarf half-light radius ($R_h$) and 
(ii) the slope of the DM halo profile at $R\approx R_h$, and is largely
independent of the initial eccentricity distribution of the binary systems 
and the anisotropy of the stellar orbits about the galactic
potential. 
For the reported velocity dispersion and half-light radius of Segue I, the closest ultra-faint, our models predict $a_t$ values that are a factor 2--3 smaller in cuspy haloes than in {\em any} of the cored models considered here. Using mock observations of Segue I we show that measuring the projected two-point correlation function of stellar pairs with sub-arcsecond resolution may provide a useful tool to constrain the amount and distribution of DM in the smallest and most DM-dominated galaxies.
\end{abstract}   


\section{Introduction} \label{sec:intro}
The distribution of mass on very small galactic scales can be used to
test a large range of dark matter (DM) models. DM made of `cold'
and collisionless particles clusters on virtually all scales, virializing into
equilibrium DM haloes whose density profiles diverge
toward the centre as $\rho\sim r^{-1}$ (e.g. Dubinski \& Carlberg
1991; Navarro, Frenk \& White 1996). In contrast, DM haloes made of
warm or self-interacting particles follow homogeneous-density profiles
with a `core' size that depends on the particle mass (Bode et
al. 2001) or cross-section (e.g. Zavala et al. 2013). Recently,
ultra-light axion-particles have attracted considerable attention as 
they can provide large cores and solve the over-abundance problem 
simultaneously (e.g. Marsh \& Pop 2015 and refs. therein). In
addition, Lombriser \& Pe\~narrubia (2015) show that cosmological
scalar fields that couple non-minimally to matter may lead to the 
inference of DM cores in dwarf spheroidal galaxies (dSphs).

Observationally, the distribution of DM in the Milky Way dSphs is a matter of
ongoing debate. Although current constraints favour cored halo models in at
least two of the brightest ($L\sim 10^7L_\odot$) dwarfs (Walker \&
Pe\~narrubia 2011), baryonic feedback may have modified the primordial
DM distribution in these systems, greatly complicating the
interpretation of inferred halo profiles (e.g. Pontzen \& Governato
2012). Even if it is becoming progressively clear that faint
($L\lesssim 10^6 L_\odot$) dSphs cannot generate sufficient feedback
energy to remove the primordial CDM cusps (Pe\~narrubia et al. 2012, di
Cintio et al. 2014), the small number of chemodynamical tracers in
these galaxies and the existence of model degeneracies introduces severe uncertainties in equilibrium
modelling techniques (Breddels \& Helmi 2013).  As a result, alternative means of constraining the DM distribution in faint dSphs are needed. 
Along this line Errani et al. (2015) showed that the internal kinematics of stellar streams associated with dSphs encode information on the progenitor's DM halo profile. 

In this paper we show that wide binary stars may offer an alternative route to probe the DM potential of the Milky Way ultra-faint ($L\lesssim
10^4L_\odot$) dSphs. 
Due to their low binding energies wide binaries are easily disrupted by the host tidal field (Heggie 1975), which we assume to be completely DM-dominated during the full dynamical history of these objects.
In \S\ref{sec:gal} we describe the equations of motion that govern the evolution of binary systems in dSphs. \S\ref{sec:nbody} outlines the $N$-body experiments, which are later used in \S\ref{sec:results} to generate mock data and explore observational signatures. The results are summarized in \S\ref{sec:summary}. 

\section{Binary systems in a galactic potential}\label{sec:gal}
\subsection{Equations of motion}\label{sec:eqmot}
Consider two stars in a binary system with masses $m_1$ and $m_2$ and a relative 
separation ${\bf r}={\bf R}_1-{\bf R}_2$ moving in a galactic
(spherical) potential $\Phi_G(R)$. In the galactic rest frame their
barycentre can be calculated as 
${\bf R}=(m_1{\bf R}_1+m_2{\bf R}_2)/m_b$, where $m_b=m_1+m_2$ is the mass of the binary system and
\begin{eqnarray}\label{eq:r}
{\bf R}_1&=&+\frac{m_2}{m_1+m_2}{\bf r}+{\bf R}\\ \nonumber
{\bf R}_2&=&-\frac{m_1}{m_1+m_2}{\bf r}+{\bf R}.
\end{eqnarray}
Hence, the equations of motion can be written as
\begin{eqnarray}\label{eq:eqmot}
\ddot{\bf R}_1&=&-\frac{G m_2}{|{\bf R}_1-{\bf R}_2|^3}({\bf R}_1-{\bf R}_2)-\nabla\Phi_G({\bf R}_1)=-\frac{G m_2}{r^3}{\bf r}-\nabla\Phi_G({\bf R}_1)\\ \nonumber
\ddot{\bf R}_2&=&-\frac{G m_1}{|{\bf R}_2-{\bf R}_1|^3}({\bf R}_2-{\bf R}_1)-\nabla\Phi_G({\bf R}_2)=+\frac{G m_1}{r^3}{\bf r}-\nabla\Phi_G({\bf R}_2).
\end{eqnarray}
In this work we shall assume that the relative separation between
the stars is much smaller than their distance to the galaxy centre,
such that $r/R\ll 1$. Inserting Equation~(\ref{eq:r})
into~(\ref{eq:eqmot}) and Taylor expanding $\nabla\Phi_G$ around the
barycentre, ${\bf R}$, yields an equation of motion for the relative
separation between the pair
\begin{eqnarray}\label{eq:eqmotsep}
\ddot{\bf r}= \ddot{\bf R}_1-\ddot{\bf R}_2\approx -\frac{G m_b}{r^3}{\bf r}+{\bf T}\cdot {\bf r};
\end{eqnarray}
where ${\bf T}$ is the {\it tidal tensor}, defined 
 \begin{eqnarray}\label{eq:tensor}
T^{ij}=-\frac{\partial^2\Phi_G}{\partial x_i\partial x_j}\bigg|_{\bf R}.
\end{eqnarray}
Since $T^{ij}$ is symmetric, it has only 6 independent components.

To compute the separation at which the mutual gravitational attraction
between the stars becomes comparable to the strength of the external
tidal field it is useful to consider the case of a binary system
moving on a circular orbit with an angular frequency 
$\Omega=v_c/R=[G M_G(<R)/R^3]^{1/2}$ about a host galaxy with a mass profile $M_G(R)$.  
Following Renaud, Gieles \& Boily (2011) we rotate the binary frame to
a new coordinate system ${\bf r}'=(x',y',z')$, where $x'$ is parallel
to {\bf R}, $y'$ is parallel to $\dot
{\bf R}$, and $z'$ is perpendicular to both $x'$ and $y'$. 
In these coordinates the effective tidal tensor $T_e=-\partial^2(\Phi_G+\Phi_c)/(\partial x_i\partial x_j)$ contains a centrifugal potential $\Phi_c=-\Omega^2r^2/2$. The eigenvalue of $T_e$ associated with the $x'$-axis ($\lambda_1$) determines
the location of the Lagrange points L1 and L2 and defines the {\it tidal radius} (Renaud
et al. 2011)
\begin{eqnarray}\label{eq:rt}
r_t\equiv \bigg[\frac{G m_b}{\lambda_1}\bigg]^{1/3},
\end{eqnarray}
where 
\begin{eqnarray}\label{eq:lambda}
\lambda_1=-\frac{\partial^2\Phi_G}{\partial x'^2}\bigg|_{{\bf R}'}+\frac{\partial^2\Phi_G}{\partial z'^2}\bigg|_{{\bf R}'}\approx \gamma\Omega^2,
\end{eqnarray}
and $\gamma(R)\equiv -\d \log \rho/\d \log R$ is the power-law slope
of the host's density profile computed at the galactocentric radius $R$. Note that in a Keplerian potential $\gamma=3$ and
$\Omega^2=GM_G/R^3$, which recovers the well-known Jacobi radius 
$r_t=R[m_b/(3M_G)]^{1/3}$.

Binary systems with separations $r\gtrsim r_t$ may be disrupted by the
host tidal field. Interestingly, Eqs.~(\ref{eq:rt})
and~(\ref{eq:lambda}) imply that the tidal radius diverges in the
limit $\gamma\to 0$, suggesting that tidal disruption will be less efficient in galaxies with shallow density 
profiles\footnote{Indeed, harmonic potentials have a {\it compressive} tidal field where all the eigenvalues become negative (see Renaud et al. for details).}.

\subsection{Dwarf spheroidal galaxies}\label{sec:rt}
Dwarf spheroidal galaxies (dSphs) are particularly interesting objects
because their gravitational potential is completely DM dominated,
i.e. $\Phi_G=\Phi_{\rm baryons}+\Phi_{\rm DM}\approx \Phi_{\rm DM}$
(e.g. Mateo 1998; Gilmore et al. 2007). Hence, measuring the relative
separation between resolved stellar pairs in these galaxies can potentially be 
used to constrain the DM potential.

For simplicity, we assume that binary systems form with a single mass
$m_b$ and initially follow the same spatial distribution as the
total stellar component, $\rho_\star(R)$, with a
half-light radius, $R_h$. Eqs.~(\ref{eq:rt})
and~(\ref{eq:lambda}) indicate that the disruption of binary systems due
to tides will be mainly determined by (i) the slope of the DM density
profile at $R\approx R_h$ and (ii) the enclosed mass
$M(<R_h)$. Several studies have shown that the latter quantity can be 
robustly inferred from observations as 
\begin{eqnarray}\label{eq:m}
M(<R_h)=\frac{5R_h\sigma_\star^2}{2G},
\end{eqnarray}
 where $\sigma_\star$ is the luminosity-averaged velocity dispersion 
(Walker et al. 2009). The mass estimate~(\ref{eq:m}) is independent of the stellar velocity anisotropy (e.g. Walker et al. 2009) and also holds in triaxial haloes drawn from cosmological simulations (Laporte et al. 2013). 

To estimate the tidal radius~(\ref{eq:rt}) for the Milky Way dSph
population we use the observed relationship between the velocity 
dispersion and half-light radius,
$\sigma_\star=\sigma_0(R_h/R_0)^\alpha$, where $\alpha\simeq 0.5$, 
$\sigma_0\simeq 0.93\kms$ and $R_0=1\pc$ (Walker et
al. 2010). Combination of Eqs.~(\ref{eq:rt}),~(\ref{eq:lambda}) 
and~(\ref{eq:m}) for $m_b=1 M_\odot$ yields
\begin{eqnarray}\label{eq:rtdsphs}
r_t= \bigg[\frac{2G m_b R_0^{2\alpha}}{5\gamma \sigma_0^2}\bigg]^{1/3}R_h^{2(1-\alpha)/3}\simeq 0.27\pc\bigg[\frac{1}{\gamma(R_h)}\cdot\frac{R_h}{10 \pc}\bigg]^{1/3},
\end{eqnarray}
which suggests that binary disruption by the smooth DM tidal field is most efficient in the smallest
dSphs. Given that the widest binary known to us has a separation of 
$\sim 1.1\pc$ (Chanam\'e \& Gould 2004; Quinn et al. 2009), galaxies with half-light radii 
$R_h\lesssim 700\pc$ provide the best targets to probe the shape of
the DM halo mass profile via the separation function of stellar pairs (see \S\ref{sec:obs}).

\section{$N$-body experiments}\label{sec:nbody}
The tidal limit estimates derived in \S\ref{sec:rt} assume that stars
move on circular orbits about the centre of the potential. In this
Section we use N-body techniques to examine more realistic orbital 
configurations that reproduce the brightness profile and velocity 
dispersion of Segue I, the closest ultra-faint dSph known to us.

\subsection{Evolution in the host galaxy potential}
We use the method of Walker \& Pe\~narrubia (2011) to generate tracer particle 
ensembles in equilibrium within a spherical DM halo potential (see
that paper for details). Briefly, we adopt a DM model 
\begin{eqnarray}\label{eq:rho}
\rho(R)=\frac{\rho_0 R_s^4}{(R+R_c)(R+R_s)^3};
\end{eqnarray}
which approaches a Hernquist (1990) profile in the limit $R_c\to 0$,
thus recovering the cuspy profile observed in self-consistent CDM 
simulations of structure formation. 

Stars are tracer particles distributed as a Plummer (1911) sphere, 
\begin{eqnarray}\label{eq:rho_star}
\rho_\star(R)=\frac{\rho_{\star,0}}{[1+(R/R_\star)^2]^{5/2}};
\end{eqnarray}
where $R_h\simeq 1.3 R_\star$. Stellar velocities are drawn from a 
Opsikov-Merrit distribution function (Osipkov 1979; Merrit
1985). These models have velocity distributions with anisotropy
profiles of the form $\beta_\star(R) \equiv 1-\overline{v^2_\theta}/\overline{v^2_r} = R^2/(R^2 +R_a^2)$,
where $R_a$ is the ``anisotropy radius''. 
Here we consider stellar models with isotropic ($R_a\to \infty$) and radially anisotropic ($R_a=R_\star$) orbital distributions.


The model parameters are chosen to reproduce the luminosity profile and
the velocity dispersion of Segue I (Belokurov et al. 2007), the
closest ultra-faint dSph ($D=23\kpc$, Martin et al. 2008). Inserting 
the (deprojected 3D) half-light radius $R_h\simeq 40\pc$ (Martin et
al. 2008; Simon et al. 2011) and the velocity dispersion
$\sigma_\star\simeq 4\kms$ (Geha et al. 2009; Simon et al. 2011) in 
Eq.~(\ref{eq:m}) leads to an enclosed mass $M(<R_h)\simeq3\times10^5M_\odot$ 
which, combined with a luminosity $L_v\sim 300 L_\odot$,
yields a mass-to-light ratio $2M(<R_h)/L\sim 2000 (M_\odot/L_\odot)$, 
making this galaxy one of the darkest known to us (Geha et
al. 2009). Note that the dynamical time of this galaxy,
$\Omega^{-1}\simeq 7\myr$, is hundreds of times shorter than the
the estimated stellar age (Belokurov et al. 2007). To find the DM
parameters $\rho_0$ and $R_s$ we impose two conditions on the halo
profile: (i) the mass enclosed within $40\pc$ is $3\times 10^5M_\odot$
and (ii) its concentration follows the relation found by 
Macci\`o et al. (2007) at redshift $z=0$. Tests with different $M-c$ 
relations yield very similar results, highlighting the fact that at
leading order the tidal field is controlled by $M(<R_h)$ and $\gamma$, 
as expected from~(\ref{eq:lambda}). We consider DM core sizes
$R_c/R_h=0,1,2$ and 3 ($R_c=0$ corresponds to a DM
cusp). With these choices the halo parameters are 
$[R_c/\pc,R_s/\pc,\rho_0/(M_\odot \pc^{-3})]=[0, 85 ,0.95]; [40,300,0.42]; [80, 913, 0.18]$ 
and $[120, 2465, 0.09]$. All our models have
scale radii $R_s\gg R_c\sim R_h$, so that the stellar populations are 
deeply embedded in the DM halo potential.

For each halo model we generate equilibrium ensembles of
$N_\star=10^4$ tracer particles. Orbits in the host potential are 
followed for $10 \gyr$ using the particle-mesh code {\sc superbox} 
(Fellhauer et al. 2000) with a time step $\Delta t=\Omega^{-1}/20$. 
For Seg I $\Omega^{-1}\simeq 7 \myr$, which yields a total 
number of time steps $N_t=10\gyr/\Delta t=28570$. At each timestep,
and for all particles, we record the 6 (independent) components of the tidal 
tensor~(\ref{eq:tensor}).

The equations of motion~(\ref{eq:eqmotsep}) are solved using a
Runge-Kutta scheme (e.g. Press et al. 1992) with a variable time-step 
chosen such that energy conservation in isolation is better than $1:1000$. To calculate the tidal force at any arbitrary time of the binary evolution we use a sync-interpolation algorithm (Pe\~narrubia et al. {\it in prep.}).


\begin{figure}
\begin{center}
\includegraphics[width=79mm]{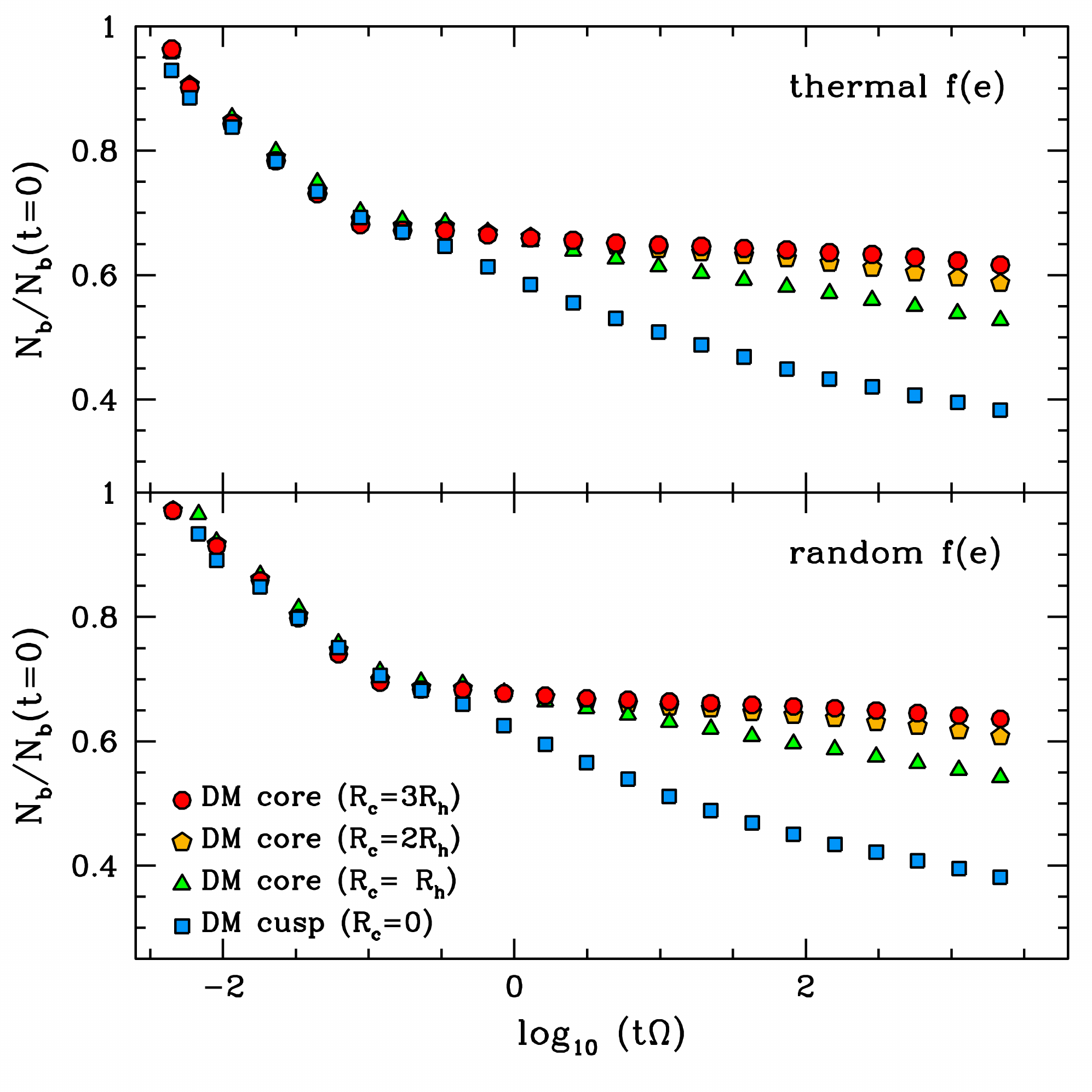}
\end{center}
\caption{Fraction of surviving binaries as a function of time. Note the two distinct evolutionary regimes, a rapid `catastrophic' evolution where tidal disruption occurs on a time-scale comparable to the orbital dynamical time of the galaxy, $t\sim \Omega^{-1}$, followed by a much slower `diffusive' evolution on time-scales $t\gg \Omega^{-1}$. Comparison between the two panels indicates no sensitivity to the initial eccentricity distribution, $f(e)$.}
\label{fig:num}
\end{figure}

\subsection{Binary systems }
We characterize the initial orbits of binary systems by their
semi-major axis distribution $q(a,t=0)$, and a eccentricity
distribution $f(e,t=0)$. Both are probability functions normalized
so that 
$\int_{a_{\rm min}}^{a_{\rm max}} q(a)da=\int_0^1 f(e)de=1$ 
at all times. 

Motivated by observations of wide binaries in the Milky Way we adopt a power-law distribution $q(a,t=0)=c_\lambda a^{-\lambda}$, 
with $c_\lambda$ being a normalization constant and 
$\lambda\gtrsim 1$. Current constraints favour a power-law index
between $\lambda= 1$ (the \"Opik 1924 distribution; 
e.g. Longhitano \& Binggeli 2010) and $\lambda\simeq1.5$ 
(e.g. Chanam\'e \& Gould 2004); for simplicity the results shown below assume $\lambda=1$, although we also test models with $\lambda=1.5$ 
Below we show that the limits $a_{\rm min}=0.02\pc$
and $a_{\rm max}=2.0\pc$ cover the relevant scales in our study. For a pair with $m_b=1M_\odot$ this implies orbital periods between $0.26$--$264\myr$.

The eccentricity distribution of binaries with $a\gtrsim 0.05\pc$
remains largely unconstrained. Recently, Tokovinin \& Kiyaeva (2016) find that nearby binaries with $a\lesssim 0.05\pc$ have a distribution $f(e)\propto \kappa e$, with $\kappa\approx 1.2$. $N$-body simulations that follow the formation of wide binaries during the dissolution of stellar clusters find $\kappa\simeq 2$ (Kouwenhoven et
al. 2010). Here we explore models with random ($f(e)={\rm const.}$) and thermal ($\kappa=2$) distributions.

\section{Results}\label{sec:results}

\subsection{Tidal disruption}
Fig.~\ref{fig:num} shows that the disruption of binary systems is less
efficient in haloes with shallow density profiles, as expected from 
the analytical estimates in \S\ref{sec:eqmot}. The binary fraction
follows two evolutionary regimes: an exponential disruption rate on a 
time-scale comparable to the orbital dynamical time of the galaxy, 
$t\sim \Omega^{-1}$, in which binaries are disrupted shortly before
or during their first pericentric approach.
This is followed by a very gentle decline on time-scales $t\gg \Omega^{-1}$
resulting from small (`diffusive') energy injections during consecutive 
pericentres. Comparison of the upper and lower panels shows that 
the initial eccentricity distribution plays a minor role in
determining the survival of wide binaries. In what
follows we focus on models with a thermal eccentricity distribution.

\begin{figure}
\begin{center}
\includegraphics[width=82mm]{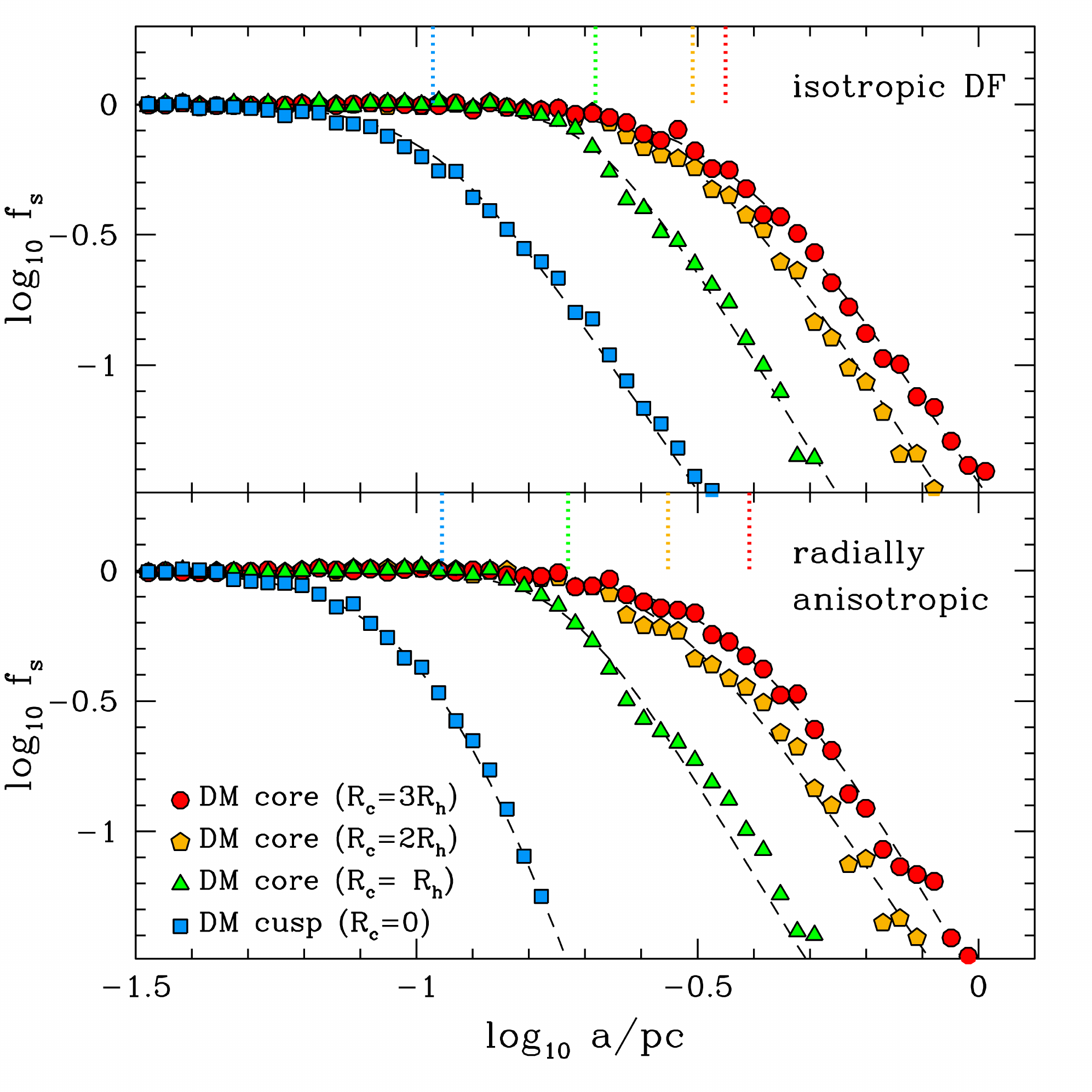}
\end{center}
\caption{Fraction of surviving binaries as a function of semi-major axis. Upper and lower panels show stellar binaries with isotropic ($R_a\to\infty$) and radially anisotropic  ($R_a=R_\star$) orbital distributions (see text). Black-dashed lines show the best-fitting functions~(\ref{eq:fit}). Vertical dotted lines mark the location of the truncation scale $a_t$.}
\label{fig:fs}
\end{figure}

Given that the binding energy of a binary system is $E=-Gm_b/(2a)$ we
expect tidal disruption to affect predominantly systems with large 
semi-major axes. This is visible in Fig.~\ref{fig:fs}, which shows
that the fraction of surviving binaries, $f_s\equiv q(a,t)/q(a,t=0)$, 
sharply drops at semi-major axes $a\gtrsim a_t$ (marked with vertical 
dotted lines for ease of reference), with a truncation $a_t$ that
shifts to larger separations as the DM core size increases relative 
to the stellar size. In contrast, binaries with $a\ll a_t$ and high binding energies survive 
regardless of the halo profile. To estimate
 $a_t$ we fit $f_s$ with a broken power-law function,
\begin{eqnarray}\label{eq:fit}
g(a)=\frac{1}{[1+(a/a_t)^m]^{n/m}},
\end{eqnarray}
which contains three free parameters: a truncation $a_t$, an
intermediate slope $m>0$ and an outer slope $n\ge 0$, such that 
$f_s\sim (a/a_t)^{-n}$ for $a\gg a_t$. For stellar binaries moving 
on isotropic orbits ($R_a\to \infty$; upper panel) the 
best-fitting parameters are $[a_t/\pc,m,n]=[0.11, 4.7, 3.2],[0.21,5.9, 3.5], [0.31, 4.1, 3.4]$, 
$[0.36, 3.8, 3.2]$ for $R_c/R_h=0, 1, 2$ 
and 3, respectively; for radially anisotropic models 
($R_a=R_\star$; lower panel) we find $[0.11, 3.8, 6.3], [0.19, 5.9,3.5], [0.28, 4.6, 3.2]$ 
and $[0.39, 3.5, 3.9]$, respectively. We checked that these values do not depend on the slope of $q(a,t=0)$ insofar as $\lambda\lesssim n$. We also find a negligible dependence to the initial eccentricity distribution, $f(e,t=0)$.
Radially anisotropic orbital distributions bring a larger fraction of binaries to the central regions of the potential. For cuspy profiles this leaves $a_t$ unchanged, but steepens the power-law index $n$. For cored models, however, the shape of the binary separation function is largely independent of the assumed velocity distribution.
The key result is that $f_s$ depends
sensitively on the DM profile: in cuspy haloes $f_s$ is truncated at 
separations a factor of 2--3 smaller than in {\em any} of the cored profiles
considered here, reflecting the weak dependence of the tidal 
radius~(\ref{eq:rt}) on galactocentric distance in regions where 
the halo slope $\gamma\to 0$.             


\begin{figure*}
\begin{center}
\includegraphics[width=160mm]{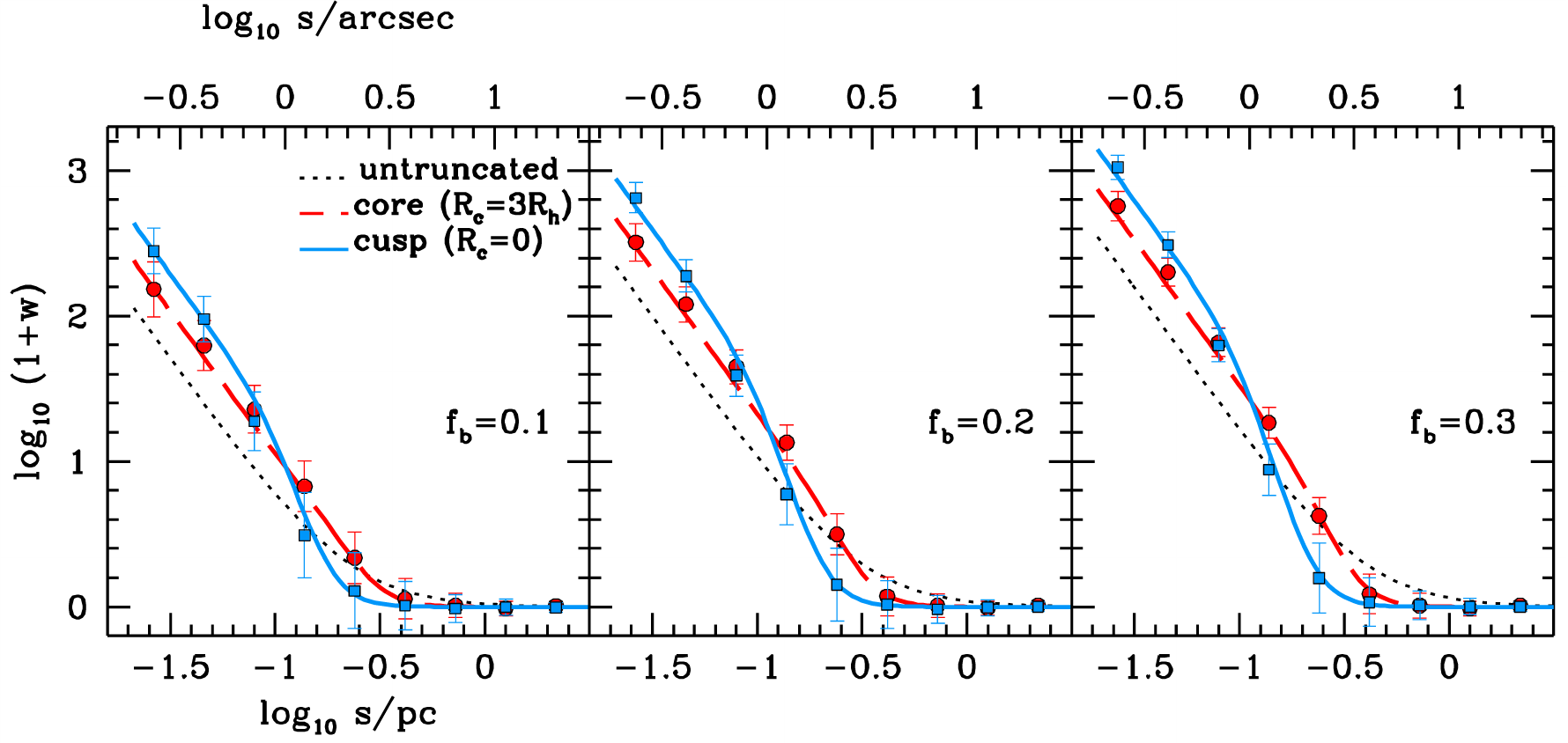}
\end{center}
\caption{Projected two point correlation function for mocks of Segue I with $N_\star=1000$ and different wide binary fractions ($f_b=N_b/N_\star$). Blue-solid and red-dashed lines correspond to the analytical estimates from Equation~(\ref{eq:psi}). Comparison with Fig.~\ref{fig:fs} shows that theoretical lines cross each other at $s\sim a_t$.}
\label{fig:2p}
\end{figure*}

\subsection{Observational signatures}\label{sec:obs}
A direct measurement of the orbital parameters of wide binaries in dSphs is
currently unfeasible owing to their long orbital periods.
To investigate whether the {\it projected} separation of stellar pairs holds 
information on the shape of the DM halo profile we construct mock
catalogues of Segue I using the isotropic $N$-body models outlined in 
\S\ref{sec:nbody}. The number of stars in this
galaxy is $N_{\star,{\rm tot}}\approx \Upsilon_\star L_v/\langle m_\star\rangle= 1500$, 
where we have assumed a stellar mass-to-light ratio
$\Upsilon_\star=2M_\odot/L_\odot$ typical for old, metal-poor stellar 
populations (Bell \& de Jong 2001), and a Kroupa (2002) IMF with a
mean stellar mass $\langle m_\star\rangle=0.4M_\odot$. We further 
assume that at most $2/3$ of $N_{\star,{\rm tot}}$ can be detected 
with current instrumentation, which yields $N_\star=1000$, of 
which $N_b=f_b N_\star$ are individual particles in binary systems with locations given by 
Eq.~(\ref{eq:r}) with $m_1=m_2$. For each mock we measure the 
projected two-point correlation function (2PCF) as
\begin{eqnarray}\label{eq:2p}
1+w(s)\equiv \frac{\psi(s)}{P(s)},
\end{eqnarray}
where $\psi(s)$ is the number of pairs with projected separations between $s,s+\d s$ and $P(s)$ is the expected number of {\it random} pairs in the same interval.  

As expected, Fig.~\ref{fig:2p} shows that presence of binaries leads
to an {\it excess} ($w>0$) of pairs at separations $s\lesssim a_t$
with respect to the random distribution. To understand the shape of
the 2PCF we can relate the function $\psi(s)$ to the
semi-major axis distribution $q(a,t)$ as (Longhitano \& Binggeli 2010)
\begin{eqnarray}\label{eq:psi}
\psi(s)\approx f_b N_\star q(\langle s\rangle,t)=f_b N_\star c'_\lambda \langle s\rangle^{-\lambda} f_s(\langle s\rangle,t),
\end{eqnarray}
where $\langle s\rangle$ is the average projected separation at a
fixed semi-major axis, $f_s\approx g$ is the survival fraction 
curves fitted in Fig.~2 (dashed lines), and $c'_\lambda$ is a 
normalization factor such that 
$\int_{\langle s\rangle _{\rm min}}^{\langle s\rangle_{\rm max}}\d s q(s,t)=1$. 
For a thermal 
eccentricity distribution $\langle s\rangle=5\pi a/16\simeq 0.98 a$ 
(Yoo et al. 2004). However, our $N$-body models indicate that the 
orbits of bianary systems with $a\gtrsim a_t$ become radially biased, decreasing 
the average separation to $\langle s\rangle/a=0.93$ and 0.70 in cuspy 
and cored ($R_c=3R_h$) halo models, respectively. Blue-solid 
and red-dashed lines in Fig.~\ref{fig:2p} show that
Eq.~(\ref{eq:psi}) provides a good match to the measured 
2PCFs. Note that at $ s\ll a_t$ we have $f_s\simeq 1$ 
and $\psi(s)\sim s^{-\lambda}$. Since the number of random pairs obeys 
$P(s)\d s\sim s\d s$, we find that $w\sim s^{-(\lambda+1)}$ at small 
separations, thus constraining the slope of the {\it unperturbed} 
semi-major axis distribution ($\lambda$). 
In addition, theoretical lines cross each other at $s\sim a_t$, and 
since $q(a,t)$ 
is normalized to unity we find that the smaller the truncation scale the larger contribution to $w$ at small separations and 
{\it vice versa}, suggesting that the 2PCF provides a useful statistical tool 
to distinguish between the halo profiles explored in
\S\ref{sec:nbody}. Whether or not a statistically-meaningful distinction is feasible with
current technology depends on the (unknown) wide binary fraction.
The Poisson error bars indicate that a robust characterization of the DM halo profile 
in galaxies with $L\lesssim 10^3L_\odot$ at a heliocentric distance $D$
requires deep astrometry data with a minimum resolution $\Delta \theta_{\rm min}\sim a_t/D$ and $f_b\gtrsim 0.1$.


\section{Discussion \& Summary}\label{sec:summary}
In this paper we use $N$-body simulations to explore the survival of wide
stellar binaries in the DM haloes of ultra-faint dSphs. We find that 
binaries with semi-major axes $a\gtrsim a_t$ are disrupted by the halo 
tidal field on time-scales $t\sim \Omega^{-1}=[GM(<R_h)/R_h^3]^{-1/2}$, 
where $M(<R_h)$ is the mass enclosed within
the half-light radius $R_h$ of the dwarf. Our models 
indicate that the truncation scale $a_t$ is sensitive to the shape of 
the DM halo mass profile. For the dynamical mass $(\simeq 3\times 10^5M_\odot)$, size ($\simeq 40\pc$) and stellar ages ($\gtrsim 10\gyr$) of Segue I, the closest 
ultra-faint, our models predict values for $a_t$ that are a factor of 2--3 smaller in cuspy haloes than in any of the cored models considered here. 

The models adopt a static, smooth DM potential in order to derive the 
dSph tidal field acting on the binary particles. Although the static 
approximation may be justified by the small evolution of DM haloes 
within the region populated by stars (e.g. Cuesta et al. 2008),
neglecting the large population of dark substructures predicted by 
$\Lambda$CDM may overestimate the survival of binaries
(Pe\~narrubia et al. 2010).  
Dense substructures (both dark and baryonic) may be particularly
damaging if they can reach the central regions of the host halo 
before being tidally disrupted (Laporte \& Pe\~narrubia 2015;
Starkenburg \& Helmi 2015; Ben{\'{\i}}tez-Llambay et al. 2016). 
Given that dynamical friction becomes inefficient in homogeneous-density
cores (Read et al. 2006) we expect encounters between stars and substructures 
to be more likely in dSphs embedded in cuspy DM haloes. Studying the impact of clumps may also be relevant for warm and axion dark matter models in which dSphs are expected to be largely devoid of dark substructures.
We plan to address these issues in future contributions by 
injecting binary populations in potentials drawn from 
cosmological hydrodynamic simulations. 

The analysis of mock data indicates that measuring the 2PCF of stellar 
pairs in the Milky Way ultra-faint dSphs may provide a useful
statistical tool to detect and characterize the semi-major 
distribution of wide binaries, and hence constrain the inner slope 
of the DM halo profile. This type of analysis requires deep
photometric data with sub-arcsecond resolution, providing a case 
for space missions like HST and the forthcoming WFIRST. 

\section{Acknowledgements}
We thank the referee F. Renaud for his insightful comments. ADL is supported by a COFUND Junior Research Fellowship. 
JC acknowledges support from Proyecto FONDECYT Regular 1130373; BASAL PFB-06 Centro de Astronom\'ia y Tecnolog\'ias Afines; and by the Chilean Ministry for the Economy, Development, and Tourism's Programa Iniciativa Cient\'ifica Milenio grant IC 120009, awarded to the Millennium Institute of Astrophysics.
MGW is supported by NSF grants AST-1313045 and AST-1412999.

{}

\end{document}